\documentclass[12pt]{article}
\usepackage{amssymb,amsmath}
\usepackage{epic,curves}
\newcommand{\intinfty}{\displaystyle\int_{-\infty}^{\infty}}
\newcommand{\intOinfty}{\displaystyle\int_{0}^{\infty}}
\newcommand{\fulldiff}[2]{\dfrac{d#1}{d#2}}
\newcommand{\partdiff}[2]{\dfrac{\partial#1}{\partial#2}}

\newcommand{\hc}{\mathrm{h.c.}}

\newsavebox{\mirrorEast}
\newsavebox{\mirrorNorth}
\newsavebox{\mirrorWest}
\newsavebox{\mirrorSouth}
\newsavebox{\beamsplSWNO}
\newsavebox{\beamsplSENW}
\newsavebox{\FabryH}
\newsavebox{\FabryV}
\newsavebox{\DelayH}
\newsavebox{\DelayV}
\newsavebox{\FabryHV}
\newsavebox{\beamsplMain}

\savebox{\mirrorEast}(2,10){
  \begin{picture}(2,10)
    \thicklines\drawline(2,0)(0,0)(0,10)(2,10)\curve(2,0,1,5,2,10)
  \end{picture}
}

\savebox{\mirrorNorth}(10,2){
  \begin{picture}(10,2)
    \thicklines\drawline(0,2)(0,0)(10,0)(10,2)\curve(0,2,5,1,10,2)
  \end{picture}
}

\savebox{\mirrorWest}(2,10){
  \begin{picture}(2,10)
    \thicklines\drawline(0,0)(2,0)(2,10)(0,10)\curve(0,0,1,5,0,10)
  \end{picture}
}

\savebox{\mirrorSouth}(10,2){
  \begin{picture}(10,2)
    \thicklines\drawline(0,0)(0,2)(10,2)(10,0)\curve(0,0,5,1,10,0)
  \end{picture}
}

\savebox{\beamsplSWNO}(10,10){
  \begin{picture}(10,10)
    \thicklines\drawline(0,0)(10,0)(10,10)(0,10)(0,0)(10,10)
  \end{picture}
}

\savebox{\beamsplSENW}(10,10){
  \begin{picture}(10,10)
    \thicklines\drawline(10,0)(10,10)(0,10)(0,0)(10,0)(0,10)
  \end{picture}
}

\savebox{\FabryH}(40,10){
  \begin{picture}(40,10)
    \put(0,0){\usebox{\mirrorEast}}
    \put(28,0){\usebox{\mirrorWest}}
    \thicklines
    \put(30,5){\vector(1,0){10}}\put(40,4){\makebox(0,0)[rt]{$x_1$}}
    \thinlines
    \put(15,5){\vector(-1,0){7}}\put(8,5){\line(-1,0){7}}
    \put(15,5){\vector(1,0){7}}\put(22,5){\line(1,0){7}}
  \end{picture}
}

\savebox{\FabryV}(10,40){
  \begin{picture}(10,40)
    \put(0,0){\usebox{\mirrorNorth}}
    \put(0,28){\usebox{\mirrorSouth}}
    \thicklines
    \put(5,30){\vector(0,1){10}}\put(6,40){\makebox(0,0)[lt]{$x_2$}}
    \thinlines
    \put(5,15){\vector(0,-1){7}}\put(5,8){\line(0,-1){7}}
    \put(5,15){\vector(0,1){7}}\put(5,22){\line(0,1){7}}
  \end{picture}
}

\savebox{\DelayH}(40,10){
  \begin{picture}(40,10)
    \put(28,0){\usebox{\mirrorWest}}
    \thicklines
    \put(2,2){\framebox(2,6)}
    \put(30,5){\vector(1,0){10}}\put(40,4){\makebox(0,0)[rt]{$x_1$}}
    \thinlines
    \drawline(0,1)(28,1)(4,5)(28,9)(0,9)
  \end{picture}
}

\savebox{\DelayV}(10,40){
  \begin{picture}(10,40)
    \put(0,28){\usebox{\mirrorSouth}}
    \thicklines
    \put(2,2){\framebox(6,2)}
    \put(5,30){\vector(0,1){10}}\put(6,40){\makebox(0,0)[lt]{$x_2$}}
    \thinlines
    \drawline(1,0)(1,28)(5,4)(9,28)(9,0)
  \end{picture}
}

\savebox{\FabryHV}(60,60){
  \begin{picture}(60,60)
    \put(20,0){\usebox{\FabryH}}
    \put(0,20){\usebox{\FabryV}}
    \thinlines
    \drawline(5,20)(5,5)(20,5)
    \curve(20,5,13,6,9,9,6,13,5,20)
  \end{picture}
}

\savebox{\beamsplMain}(45,45){
  \begin{picture}(45,45)
    \put(20,20){\usebox{\beamsplSWNO}}
    \put(20,20){\makebox(0,0)[rt]{\sf BS}}
    \thicklines
    \put(20,35){\line(1,1){10}}\put(25,40){\makebox(0,0)[rb]{\sf M1}}
    \put(35,20){\line(1,1){10}}\put(40,25){\makebox(0,0)[lt]{\sf M2}}
    \thinlines
    \drawline(25,15)(25,40)(40,40)(40,25)(15,25)
    \put(10,25){\vector(1,0){5}}
    \put(10,25){\vector(-1,0){5}}\put(5,25){\line(-1,0){5}}
    \put(25,15){\vector(0,-1){5}}\put(25,10){\line(0,-1){5}}
    \put(0,26){\makebox(0,0)[lb]{Light port}}
    \put(26,5){\makebox(0,0)[lb]{Dark port}}
  \end{picture}
}

\title{Quantum speedmeter and laser interferometric gravitational-wave
antennae}

\author{F.Ya.Khalili}

\date{}

\begin{document}

\maketitle

\begin{abstract}

A new topology of laser interferometric gravitational-wave antenna is
considered. It is based on two schemes: {\em quantum speedmeter} and {\em
zero-area Sagnac interferometer} and allows to obtain sensitivity better than
the Standard Quantum Limit in wide band without any large-scale modifications
of the standard topology of the laser interferometric antennae.

\end{abstract}

\section{Introduction}

It is well known that the only variables that can be continuously monitored
with arbitrary high precision are the Quantum Non-Demolition (QND) observables
that are the variables which operators (in the Heisenberg picture) commute
with themselves at different moments of time \cite{77a1eBrKhVo, Thorne1978,
92BookBrKh}. A free mass has two such variables: momentum $p$ and variable
with explicit time dependence $x-pt/m$. In both cases the meter must be able
to ``see'' momentum of free mass. However, implementation of such meters at
the quantum level of sensitivity  does not seem possible for the contemporary
level of technology.

In the article \cite{90a1BrKh} it was proposed to measure velocity of a free
test mass instead of momentum. Velocity is not a true QND observable, and it
is perturbed by the meter during such a measurement. However, its properties
are close to properties of the momentum and therefore the perturbation can be
rather easily excluded from the output signal of the meter using
cross-correlation between the measurement noise and back-action noise (see
details in the next section). It was shown also in the article \cite{90a1BrKh}
that velocity can be measured using pairs of position measurements separated
by small time $\tau$. The main requirement for such a procedure is that
decoherence of the quantum state of the meter between these two measurements
must be sufficiently small:

\begin{equation}\label{decoh}
  \tau_{\rm decoh} \gg \Omega_{\rm signal}^{-1} \,,
\end{equation}
where $\tau_{\rm decoh}$ is the decoherence time and $\Omega_{\rm signal}$ is
the characteristic frequency of the signal.

In the article \cite{90a1BrKh} two schemes of velocity measurement were
proposed: a simple semi-gedanken scheme close to the one considered in the
next section and the scheme based on two coupled microwave cavities. In the
article \cite{98a1BrGoKh} it was proposed to use the second one as a local
meter in the intracavity topology of laser gravitational-wave antennae and it
was shown that it allows to obtain sensitivity better than the Standard
Quantum Limit (SQL) using significantly lower amount of optical pumping energy
than in traditional (extracavity) schemes.

In the article \cite{00a1BrGoKhTh} the last scheme was analyzed in detail. It
was shown that it is feasible using current technology to beat the SQL by the
factor of $\sim 2$ using high-quality cryogenic microwave resonators. The main
limiting factor here is the microwave resonator relaxation time achievable in
the real experiment, which must be large due to the condition (\ref{decoh}).
It is necessary to note that eigenfrequency of one of the cavities have to be
dependent on the position of the test mass, which limits greatly achievable
quality factors of the microwave resonators (see brief analysis in the section
V of the article \cite{00a1BrGoKhTh}).

In that article it was also proposed the adaptation of this scheme into the
optical band, which can be used directly in the extracavity topology of the
gravitational-wave antennae and which also allows to obtain sensitivity better
than the SQL. It is based on the fact that using the best modern
high-reflectivity mirrors \cite{Kimble1992}, the values as long as $1\div
10\,{\rm s}$ can be obtained for the Fabry-Perot cavities with length
$L\gtrsim 10^5\,{\rm cm}$ typical for the interferometric gravitational-wave
antennae. Therefore, condition (\ref{decoh}) can be fulfilled relatively
easyly for the typical frequencies of the gravitational-wave signal
$\Omega_{\rm signal} \sim 10^2\div 10^3\,{\rm s}^{-1}$.

This idea was developed in the articles \cite{Purdue2001, PurdueChen2002}. In
these articles different topologies and regimes of the quantum speedmeter
scheme based on large-scale Fabry-Perot cavities was analyzed in detail.
It was shown that sensitivity several times better than the SQL can be
achieved at the current technological level.

However, these schemes has serious disadvantages: they require huge value of
the pumping power or an additional large-scale Fabry-Perot cavity.

Another topology of the interferometer, which is sensitive to velocity
instead of the test masses position is a zero-area Sagnac interferometer
\cite{Byer1996}. Prototypes of such a device have been built and the
characteristic frequency response proportional to the observation frequency
$\Omega$ was demonstrated \cite{Byer1999, Byer2000}. In the article
\cite{Chen2002} a quantum-mechanical study of Sagnac interferometers was
carried out.

Traditionally in zero-area Sagnac interferometers optical delay lines or ring
cavities are used instead of Fabry-Perot cavities, and it is these topologies
was considered in the paper \cite{Chen2002}. At the same time, Fabry-Perot
cavities are more convenient for the kilometer-scale interferometers from the
technological point of view.

In this paper the new version of the quantum speedmeter which combines the
zero-area Sagnac interferometer with Fabry-Perot cavities is presented. Only
two such cavities are used and therefore only minimal changes in the standard
Fabry-Perot/Michelson topology of the laser gravitational-wave antennae are
required in the proposed scheme.

In the section \ref{sec:general} general properties of all quantum speedmeter
schemes are considered. The main goal of this section is to show, that the
mentioned above cross-correlation between the measurement noise and
back-action noise is necessary in order to evade SQL in this scheme due to the
subtle difference between momentum and velocity of a free mass.

In the section \ref{sec:extra} the Sagnac/Fabry-Perot quantum speedmeter
scheme is analyzed and the sensitivity of this scheme which can be obtained
using contemporary technologies is estimated.

\section{General properties of the Quantum Speedmeter}\label{sec:general}

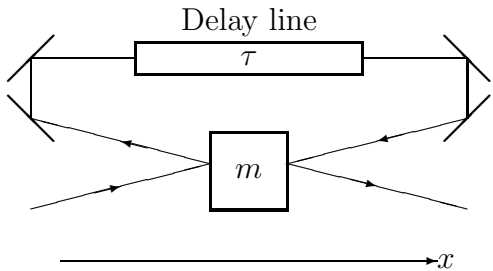
\begin{figure}

\begin{center}

\begin{picture}(74,40)

\thicklines

\put(22,30){\framebox(30,4)[cc]{$\tau$}}
\put(22,35){\makebox(30,10)[cb]{Delay line}}

\put(32,12){\framebox(10,10)[cc]{$m$}}

\put(5,27){\line(1,-1){6}}
\put(5,29){\line(1,1){6}}
\put(69,29){\line(-1,1){6}}
\put(69,27){\line(-1,-1){6}}

\thinlines

\put(8,12){\vector(4,1){12}}\put(20,15){\line(4,1){12}}
\put(32,18){\vector(-4,1){12}}\put(20,21){\line(-4,1){12}}

\drawline(8,24)(8,32)(22,32)\drawline(52,32)(66,32)(66,24)

\put(66,24){\vector(-4,-1){12}}\put(54,21){\line(-4,-1){12}}
\put(42,18){\vector(4,-1){12}}\put(54,15){\line(4,-1){12}}

\put(12,5){\vector(1,0){50}}
\put(62,5){\makebox(0,0)[lc]{$x$}}

\end{picture}

\end{center}

\caption{Gedanken scheme of the quantum speedmeter}\label{fig:simplest}

\end{figure}

\subsection{The simplest gedanken scheme}\label{sec:gedanken}

Simple gedanken scheme illustrating the idea and the main properties of the
quantum speedmeter is presented in Fig.\,\ref{fig:simplest}.

Here an optical pulse is reflected from the left side of the test mass $m$,
then passes through the delay line and then is reflected again from the other
side of the test mass. If test mass does not move then the total phase shift
of the optical pulse does not depend on its initial position. However if it
does move then there will be some phase shift

\begin{equation}
  \delta\varphi = \frac{2\omega_o\bar v\tau}{c} \,,
\end{equation}
where $\tau$ is the time interval between two reflections and $\bar v$ is the
velocity of the test mass during this interval. Therefore, monitoring the
phase of the output light beam it is possible to measure the velocity $\bar v$
with precision

\begin{equation}\label{ged_meas}
  \Delta \bar v_{\rm meas} = \frac{c\Delta\varphi}{2\omega_o\tau} \,,
\end{equation}
where $\Delta\varphi$ is the phase uncertainty of the optical pulse.

During this procedure the test mass obtains two equal kicks in opposite
directions. Therefore, its velocity {\it after} the measurement is not
perturbed. However, velocity {\it between} the kicks is perturbed by the value

\begin{equation}\label{ged_pert_v}
  \Delta v_{\rm pert} = \frac{2\Delta{\cal E}}{mc} \,,
\end{equation}
where $\Delta{\cal E}$ is the energy uncertainty of the light pulse.

From this follow two conclusions. First, position of the test mass after the
measurement is perturbed by the value

\begin{equation}\label{ged_pert_x}
  \Delta x_{\rm pert} = \frac{2\Delta{\cal E}\tau}{mc} \,.
\end{equation}
It can be obtained from the formulas (\ref{ged_meas}) and (\ref{ged_pert_x})
that

\begin{equation}
  \Delta \bar v_{\rm meas}\Delta x_{\rm pert} =
  \frac{\Delta{\cal E}\Delta\varphi}{2m\omega_o} \ge \frac{\hbar}{2m} \,,
\end{equation}
in accord with the uncertainty relation.

Second, measurement error for the {\em initial} value of the velocity of the
test mass contains term (\ref{ged_pert_v}) in addition to (\ref{ged_meas}).
Simple optimization of these two terms gives the value which is exactly equal
to the SQL:

\begin{equation}\label{delta_v_SQL}
  \Delta v_{\rm SQL} = \sqrt{\frac{\hbar}{m\tau}} \,.
\end{equation}
In the article \cite{90a1BrKh} it was shown, however, that this additional
error can be eliminated by using cross-correlation between the measurement
error and perturbation. In the scheme described above this cross-correlation
can be obtained either by using non-classical quantum state of the light pulse
with correlated phase and energy, or by detecting the output light beam by
homodyne detector with optimally chosen phase of the local oscillator. In both
cases the measurement error for the initial value of velocity will be equal to

\begin{equation}
  \Delta v_{\rm opt} = \frac{\hbar c}{2\tau\Delta{\cal E}} \,.
\end{equation}

\subsection{Single velocity measurement}

The general abstract scheme of the quantum speedmeter represents the direct
generalization of the example considered above. Let again $m$ be a free mass
which velocity $v$ have to be measured. Let ${\cal N}$ be the observable of
the meter, which provide its coupling with the test mass, and $\alpha(t)$ is
the coupling factor, $\alpha(t)=1$ when the meter is on and $\alpha(t)=0$ when
the meter is off. Suppose also that we can neglect the variable ${\cal N}$
self evolution during the measurement (this is a reasonable assumption because
in real schemes of the speedmeter, and in the gedanken example considered
above, the number of quanta in e.m. wave corresponds to this variable). Due to
this assumption, we can set the Hamiltonian of the meter equal to zero (see
also the article \cite{02a1BrGoKhMaThVy}).

Lagrangian of this system can be presented as

\begin{equation}
  {\cal L} = \frac{mv^2}{2}  - \alpha(t)v{\cal N} \,,
\end{equation}
Therefore, momentum of the mass $m$ is equal to

\begin{equation}
  p = \partdiff{\cal L}{v} = mv - \alpha(t){\cal N} \,,
\end{equation}
and the Hamiltonian of the system is equal to

\begin{equation}
  {\cal H} = pv - {\cal L} = \frac{[p+\alpha(t){\cal N}]^2}{2m}\,.
\end{equation}

The output signal of the meter is represented by observable $\Phi$ canonically
conjugated to ${\cal N}$ (if ${\cal N}$ corresponds to he number of quanta in
e.m. wave then $\Phi$ corresponds to the phase of this wave). Evolution of
the test object position $x$ and the observable $\Phi$ is described by the
equations

\begin{align}
  \fulldiff{\hat x(t)}{t} &\equiv \hat v(t)
    = \frac{\hat p+\alpha(t)\hat{\cal N}}{m} \,, \label{single_eq_a} \\
  \fulldiff{\hat\Phi(t)}{t} &= \alpha(t)\hat v(t) \,.
\end{align}
Solution of these equations can be presented in the following form:

\begin{align}
  \hat x(\tau) &= \hat x(0) +\frac{\hat p\tau}{m}
    + \frac{\hat{\cal N}\tau}{m} \,, \label{single_soln_a} \\
  \hat\Phi(\tau) &= \hat\Phi(0) + \hat{\bar v}\tau
   = \hat \Phi(0) + \frac{\hat{\cal N}\tau}{m} + \hat v_{\rm init}\tau\,,
  \label{single_soln_b}
\end{align}
where $\bar v = (p+{\cal N})/m$ is the value of velocity when the meter is turned
on, $v_{\rm init} = p/m$ is the initial value of the velocity, and $\tau$ is
the duration of the measurement.

Therefore, the momentum of the test mass is not perturbed and its velocity is
perturbed during the measurement by the value of $\alpha(t){\cal N}/m$, but
returns to the initial value after the end of the measurement [see formula
(\ref{single_eq_a})]. Perturbation of the position of the test mass after the
measurement is proportional to the uncertainty of observable ${\cal N}$ [see
formula (\ref{single_soln_a})]:

\begin{equation}
  \Delta x_{\rm pert} = \frac{\Delta{\cal N}\tau}{m} \,.
\end{equation}

By measuring variable $\Phi$ after the interaction of the meter with the test
object one can estimate the {\em perturbed} value of velocity $\bar v$ with
precision [see formula (\ref{single_soln_b})]

\begin{equation}
  \Delta\bar v_{\rm meas} = \frac{\Delta\Phi}{\tau} \,.
\end{equation}
It follows from the last two formulas, that

\begin{equation}
  \Delta x_{\rm pert}\Delta\bar v_{\rm meas}
    = \frac{\Delta\Phi\Delta{\cal N}}{m} \ge \frac{\hbar}{2m} \,.
\end{equation}
At the same time, the measurement error for the {\em initial} unperturbed
value of the velocity $v_{\rm init} = p/m$ depends on both $\Delta\Phi$ and
$\Delta{\cal N}$ and ``naive'' optimization gives that the precision is
limited by the value (\ref{delta_v_SQL}).

However, using cross-correlation between the measurement error and back-action
it is possible to measure $v_{\rm init}$ with arbitrary high precision.
Similar to the previous example, this cross-correlation can be implemented
whether by preparation of the meter in the initial state with well-defined
value of the combination $\Phi(0)+{\cal N}\tau/m$, or by monitoring the
variable $\Phi(\tau)-{\cal N}\tau/m$ instead of $\Phi(\tau)$. In both cases it
is possible to measure the initial velocity with precision

\begin{equation}
  \Delta v_{\rm opt} = \frac{\hbar}{2\tau\Delta{\cal N}} \,.
\end{equation}

\subsection{Continuous monitoring of the velocity}\label{sec:cont}

Convenient method for consideration the continuous monitoring of the velocity
of the quantum test body is based on {\em linear quantum meter} approach
\cite{92BookBrKh, 02a1BrGoKhMaThVy}. Let a classical signal force $F_{\rm
signal}$ which has to be detected acts on a quantum test mass $m$. Suppose
that the velocity of test mass $v(t)$ is monitored by the quantum speed meter.
Output signal of such a meter is proportional to a sum of the current value of
velocity and the measurement noise  $v_{\rm fluct}(t)$:

\begin{equation}\label{cont_tilde_v}
  \tilde v(t) = \hat v(t) + \hat v_{\rm fluct}(t) \,,
\end{equation}
At the same time, the meter perturbs the test mass by means of the random
force $F_{\rm fluct}(t)$. In the case of speedmeter, this force have not to
perturb momentum of the test object. It can be shown that in this case the
force can be presented in the form

\begin{equation}
  \hat F_{\rm fluct}(t) = \fulldiff{\hat p_{\rm fluct}(t)}{t} \,,
\end{equation}
where $\hat p_{\rm fluct}(t)$ is some another random function. Therefore,

\begin{equation}\label{cont_v}
  \hat v(t) = \hat v_{\rm init} + \frac{1}{m}
    \left(\int_0^t F_{\rm signal}(t')\,dt + \hat p_{\rm fluct}(t)\right) \,.
\end{equation}
It follows from the equations (\ref{cont_tilde_v}) and (\ref{cont_v}) that
signal to noise ratio for such system is equal to:

\begin{equation}
  \frac{s}{n}
  = \intinfty\frac{|F_{\rm signal}(\Omega)|^2}{S_{\rm total}(\Omega)}\,
      \frac{d\Omega}{2\pi} \,,
\end{equation}
where $\Omega$ is the observation frequency and

\begin{equation}
  S_{\rm total}(\Omega) = \Omega^2(m^2 S_v + 2mS_{vp} + S_p) \,,
\end{equation}
is the spectral density of the total net noise of the meter. Here $S_v$ and
$S_p$ are the spectral densities of the noises $v_{\rm fluct}(t)$ and $p_{\rm
fluct}(t)$ correspondingly, and $S_{vp}$ is their cross spectral density,
which satisfy the uncertainty relation

\begin{equation}
  S_vS_p - S_{vp}^2 \ge \frac{\hbar^2}{4} \,.
\end{equation}
Suppose first that $S_{vp}=0$ (there is no cross-correlation). It is easy to
show that in this case

\begin{equation}
  S_{\rm total}(\Omega) \ge \hbar m\Omega^2
\end{equation}
for all possible values of $S_v$ and $S_p$. This value exactly corresponds to
the spectral form of the SQL \cite{00a1BrGoKhTh}.

However, if $S_{vp}$ is chosen in the optimal way:

\begin{equation}\label{cont_opt_corr}
  S_{vp} = -\frac{S_p}{m} \,,
\end{equation}
then the total noise is limited only by the condition

\begin{equation}\label{cont_opt}
  S_{\rm total}(\Omega) \ge \frac{\hbar^2\Omega^2m^2}{4S_p}\,,
\end{equation}
and can be made arbitrary small if sufficient value of $S_p$ is provided.

It is important that the cross-correlation (\ref{cont_opt_corr}) does not
depend on the observation frequency. Therefore it can be implemented easily
using homodyne detector with fixed value of the local oscillator phase. The
variation measurement \cite{98a1Vy, 02a1KiLeMaThVy} also allows to obtain
sensitivity better than the SQL using the back-action and measurement noises
cross-correlation. However, in the latter case sophisticated time- or
frequency-dependence of the local oscillator phase is required in order to
overcome the SQL in the wide band.

\section{The optical speedmeter}\label{sec:extra}

\subsection{The idea of the optical speedmeter}

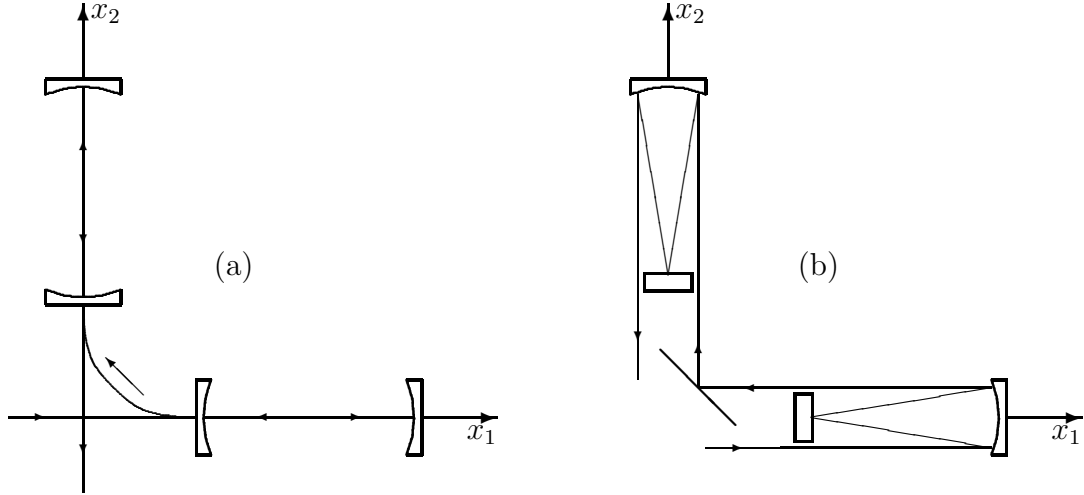
\begin{figure}

\begin{center}

\begin{picture}(65,65)

\thicklines

\put(5,5){\usebox{\FabryHV}}

\thinlines

\put(0,10){\vector(1,0){5}}\put(5,10){\line(1,0){5}}
\put(18,13){\vector(-1,1){5}}
\put(10,10){\vector(0,-1){5}}\put(10,5){\line(0,-1){5}}

\put(30,30){\makebox(0,0)[cc]{(a)}}

\end{picture}
\hspace{1cm}
\begin{picture}(65,65)

\put(25,5){\usebox{\DelayH}}
\put(5,25){\usebox{\DelayV}}

\thicklines\put(9,19){\line(1,-1){10}}

\thinlines

\put(15,6){\vector(1,0){5}}\put(20,6){\line(1,0){5}}
\put(25,14){\vector(-1,0){5}}\put(20,14){\line(-1,0){6}}
\put(14,14){\vector(0,1){6}}\put(14,20){\line(0,1){5}}
\put(6,25){\vector(0,-1){5}}\put(6,20){\line(0,-1){5}}

\put(30,30){\makebox(0,0)[cc]{(b)}}

\end{picture}

\end{center}

\caption{The idea of the optical speedmeter}\label{fig:extra1}

\end{figure}

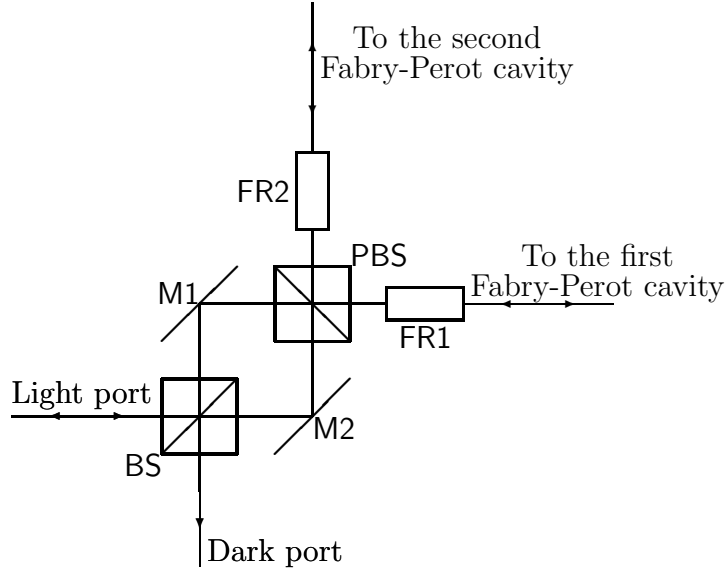
\begin{figure}

\begin{center}

\begin{picture}(90,80)

\put(0,0){\usebox{\beamsplMain}}
\put(35,35){\usebox{\beamsplSENW}}\put(45,45){\makebox(0,0)[lb]{\sf PBS}}

\thicklines

\put(50,38){\framebox(10,4)}\put(55,37){\makebox(0,0)[ct]{\sf FR1}}

\put(38,50){\framebox(4,10)}\put(37,55){\makebox(0,0)[rc]{\sf FR2}}

\thinlines

\drawline(40,50)(40,40)(50,40)

\put(10,25){\vector(1,0){5}}
\put(10,25){\vector(-1,0){5}}\put(5,25){\line(-1,0){5}}
\put(25,15){\vector(0,-1){5}}\put(25,10){\line(0,-1){5}}

\put(70,40){\vector(-1,0){5}}\put(65,40){\line(-1,0){5}}
\put(70,40){\vector(1,0){5}}\put(75,40){\line(1,0){5}}

\put(40,70){\vector(0,-1){5}}\put(40,65){\line(0,-1){5}}
\put(40,70){\vector(0,1){5}}\put(40,75){\line(0,1){5}}

\put(0,26){\makebox(0,0)[lb]{Light port}}

\put(26,5){\makebox(0,0)[lb]{Dark port}}

\put(61,41){\shortstack{To the first \\ Fabry-Perot cavity}}

\put(41,70){\shortstack{To the second \\ Fabry-Perot cavity}}

\end{picture}

\end{center}

\caption{A possible design of the central station of the optical
speedmeter}\label{fig:extra3}

\end{figure}

Suppose that we are managed, by some means, to separate output light beams
from the input ones for both of the Fabry-Perot cavities used in the standard
topology of the laser gravitational-wave antennae. One of the possible designs
which allows to do it will be presented below (see Fig.\,\ref{fig:extra3}). In
this case it is possible to direct the light beam first to the one of the
Fabry-Perot cavities and then to the second one as it is shown in
Fig.\,\ref{fig:extra1}(a). In principle, delay lines can be used in this
scheme too and they allow to create ``round-robin'' passage of the pumping
beam through the both arms more easily [see Fig.\,\ref{fig:extra1}(b)], as it
was proposed in the article \cite{Chen2002}. However, Fabry-Perot cavities are
more convenient for the kilometer-scale interferometers from the technological
point of view. Therefore, Fabry-Perot cavities will be considered in this
article.

Using rather crude but qualitatively correct approximation, phase of the
output beam can be presented in the following form (the rigorous analysis is
provided in the Appendix\,\ref{app:extra}):

\begin{equation}
  \varphi_{\rm out}(t) = \varphi_{\rm in}(t) +
    \frac{2\omega_o\tau^*}{L}\bigl[2\bar x_+(t) + \bar v_-(t)\tau^*)\bigr] \,,
\end{equation}
where

\begin{align}
  \bar x_+(t) &= \frac{x_+(t)+x_+(t-\tau^*)}{2} \,, \nonumber \\
  \bar v_-(t) &= \frac{x_-(t)-x_-(t-\tau^*)}{\tau^*} \,,
\end{align}
and

\begin{align}
  x_+ &= \frac{x_1+x_1}{2} \,, & x_- &= \frac{x_1-x_2}{2} \,.
\end{align}
Therefore, this scheme provides information about the velocity of the
differential (anti-symmetric) motion of the end mirrors and about the position
$x_+$ which corresponds to the symmetric motion.

In the gravitational-wave antennae the anti-symmetric mode is coupled with the
gravitational-wave signal. In order to eliminate information about the
position $x_+$ (the presence of this information, as it can be easily shown,
does not permit to obtain sensitivity better than the SQL), two optical beams
have to be used. One of them must ``visit'' the Fabry-Perot cavities in the
clockwise order, as it is shown in Fig.\,\ref{fig:extra1} and the second one
in the counter-clockwise order. Then the phase difference of these beams has
to be measured. This modification allows also to protect the scheme from the
frequency instability of the input laser beam.

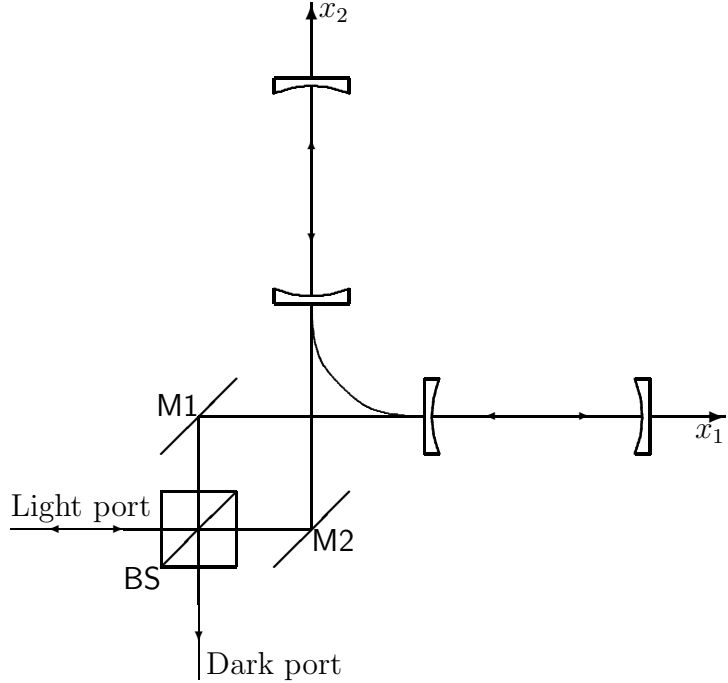
\begin{figure}

\begin{center}

\begin{picture}(95,95)

\put(0,0){\usebox{\beamsplMain}}

\put(35,35){\usebox{\FabryHV}}

\end{picture}

\end{center}

\caption{The conceptual design of the optical speedmeter}\label{fig:extra2}

\end{figure}

The measurement of the phase difference of these two beams can be performed
exactly in the same way as in the standard topology of the gravitational-wave
antennae (see Fig.\,\ref{fig:extra2}). Here beam from the pumping laser is
divided into two beams by the beam splitter. One of these beams leaves the
beam splitter from the top, is reflected from the auxiliary mirror {\sf M1},
from the right and then the upper Fabry-Perot cavities, and returns to the
beam splitter from the right side. The second beam goes in the opposite
direction. If the end mirrors of the Fabry-Perot cavities do not move then
both beams gain equal phase shifts and all optical power returns to the
pumping laser. However, the differential motion of the end mirrors will
produce differential phase shift in both beams, proportional to the speed of
this motion. In this case the part of pumping power will be splitted to the
dark port and registered by the detector.

One of the possible designs of the central station for the optical speedmeter
scheme is presented in Fig.\,\ref{fig:extra3}.\footnote{This scheme should be
considered as an example only; it is very possible that better implementations
of the idea presented here are exist. In particular, quarter-wave plates can
be used instead of the Faraday rotators and, correspondingly, the circular
polarization instead of the linear one \cite{Shanti}.} Here {\sf BS} is the
ordinary 50\%/50\ beam splitter, {\sf PBS} is the polarization beam splitter,
and {\sf FR1}, {\sf FR2} are the Faraday rotators which rotate polarization of
the light which pass through them by $45^\circ$.

Suppose that {\sf PBS} reflects the light with vertical ($90^\circ$)
polarization and is transparent for the light with horizontal ($0^\circ$)
polarization. In this case the pumping laser has to generate horizontally
polarized beam.

Consider, for example, the beam which goes from the {\sf BS} to the upper
direction. It is reflected from the auxiliary mirror {\sf M1}, passes through
the {\sf PBS} and then through the {\sf FR1}, and goes to the first
Fabry-Perot cavity having $45^\circ$ polarization. After reflection from the
Fabry-Perot cavity, it passes through the {\sf FR1} for the second time. At
this moment it has vertical polarization and hence is reflected from {\sf PBS}
and goes to the upper direction. Then it passes through {\sf FR2}, is
reflected from the second Fabry-Perot cavity, passes through {\sf FR2} for the
second time (now it has horizontal polarization again), passes through the
{\sf PBS}, is reflected from the second auxiliary mirror {\sf M2}, and returns
to the main beam splitter {\sf BS} from the right direction.

\subsection{Sensitivity of the optical speedmeter}\label{sec:extra2}

In the Appendix \ref{app:extra} the explicit values of the spectral densities
introduced in the section \ref{sec:cont} are calculated for the scheme of the
optical speedmeter described above. Here we consider for simplicity only the
particular case when observation frequency $\Omega$ small compared to the
half-bandwidth of the Fabry-Perot cavities $\gamma$. In this case, these
spectral densities are equal to [see the formulas (\ref{S_v}, \ref{S_p},
\ref{S_vp})]:

\begin{align}
  S_v &= \frac{\hbar L^2\gamma^4}{64\omega_ow\cos^2\varphi} \,, \\
  S_p &= \frac{16\hbar\omega_ow}{L^2\gamma^4} \,, \\
  S_{vp} &= \frac{\hbar}{2}\tan\varphi \,,
\end{align}
where $w$ is the pumping power, and $\varphi$ is the phase shift between
the pumping beam and the local oscillator beam.

It follows from the conditions (\ref{cont_opt_corr}) and (\ref{cont_opt}) that
the optimal value of $\varphi$ is equal to

\begin{equation}
  \varphi = -\arctan\frac{32\omega_ow}{mL^2\gamma^4} \,,
\end{equation}
and corresponding optimal spectral density of the total noise is described by
the formula

\begin{equation}\label{xi_2_extra}
  \xi^2 \equiv \frac{S_{\rm total}(\Omega)}{S_{\rm SQL}(\Omega)}
  = \frac{mL^2\gamma^4}{64\omega_o w} \,.
\end{equation}
where

\begin{equation}
  S_{\rm SQL}(\Omega) = \hbar m\Omega^2
\end{equation}
is the spectral density of the total noise which corresponds to the SQL
\cite{00a1BrGoKhTh}.

Formula (\ref{xi_2_extra}) has the structure typical for all interferometric
meters (see, for example, articles \cite{98a1Vy, 02a1KiLeMaThVy, 00a1DaKhVy}),
which originates from the Energetic Quantum Limit \cite{92BookBrKh,
00p1BrGoKhTh}. Due to this limitation, in order to obtain the sensitivity
equal to the SQL ($\xi=1$), the pumping power has to be close to the value
planned for the contemporary large-scale gravitational-wave antennae and in
order to obtain better sensitivity it is necessary to increase pumping power
proportionally to $\xi^{-2}$.

Consider, for example, values of the parameters typical for the
gravitational-wave antenna LIGO-I: $m=10\,{\rm Kg}$, $L=4\,{\rm Km}$,
$\gamma=10^3\,{\rm s}^{-1}$ and $\omega_o=2\cdot 10^{15}\,{\rm s}^{-1}$. In
this case

\begin{equation}\label{xi2}
  \xi^2 \approx \frac{1.25\,{\rm KWt}}{w} \,.
\end{equation}
It have to be noted, that this value of the pumping power can be reduced using
the power recycling, exactly as it is planned for the standard topology of the
LIGO. However, power circulating {\it inside} the interferometer remains the
same. Formula (\ref{xi2}) in this case describes the power passing through the
beam splitter.

\section{Conclusion}

Sensitivity of the scheme of the optical speedmeter considered here is typical
for schemes with traditional (extracavity) topology. If pumping beam in
coherent quantum state is used then it is necessary about one kilowatt
(without power recycling) of the optical power in order to reach the Standard
Quantum Limit, and the necessary pumping power depends on the required
sensitivity as $h^{-2}$, where $h$ is the signal amplitude which has to be
detected.

However, this scheme has the following advantages in comparison with other
proposed schemes with extracavity topology (see, for example, references
\cite{00a1BrGoKhTh, 98a1Vy, 02a1KiLeMaThVy, 01a2Kh, Buonanno2002}):

\begin{itemize}
  \item It allows to obtain sensitivity better than the Standard Quantum Limit
    in wide band (the bandwidth is limited by the bandwidth of the Fabry-Perot
    cavities only).

  \item It does not require exact information about the shape and arrival time
    of the signal.

  \item It does not require any large-scale modifications of the standard
    topology of the laser interferometric antennae.

  \item It does not require non-classical state of the pumping power
    (however, using squeezed quantum state it is possible to reduce the value
    of the pumping power)
\end{itemize}

Therefore, considered scheme looks as the promising option for the first step
beyond the Standard Quantum Limit.

\section*{Acknowledgments}

Author thanks V.B.Braginsky and S.L.Danilishin for useful remarks.

This paper was supported in part by US National Science Foundation, by the
Russian Foundation for Basic Research, and by the Russian Ministry of Industry
and Science.

\appendix

\section{Spectral densities of the noises of the optical speedmeter}
\label{app:extra}

\subsection{Intermediate formulas}

\subsubsection{Single moving mirror}\label{sec:MM}

\begin{figure}

\begin{center}

\begin{picture}(40,10)

\thicklines

\put(20,0){\usebox{\mirrorWest}}

\thinlines

\put(0,6){\vector(1,0){10}\line(1,0){10}}
\put(19,7){\makebox(0,0)[rb]{$a$}}

\put(20,4){\vector(-1,0){10}\line(-1,0){10}}
\put(19,3){\makebox(0,0)[rt]{$b$}}

\put(22,5){\vector(1,0){10}\makebox(0,0)[lc]{$x$}}

\end{picture}

(a)

\begin{picture}(70,25)

\thicklines

\put(20,5){\usebox{\mirrorEast}}
\put(20,16){\makebox(0,0)[cb]{$-R,iT$}}

\put(50,5){\usebox{\mirrorWest}}

\thinlines

\put(0,11){\vector(1,0){10}\line(1,0){10}}
\put(19,12){\makebox(0,0)[rb]{$a$}}

\put(20,9){\vector(-1,0){10}\line(-1,0){10}}
\put(19,8){\makebox(0,0)[rt]{$b$}}

\put(22,11){\vector(1,0){14}\line(1,0){14}}
\put(49,12){\makebox(0,0)[rb]{$a'$}}

\put(50,9){\vector(-1,0){14}\line(-1,0){14}}
\put(49,8){\makebox(0,0)[rt]{$b'$}}

\put(52,10){\vector(1,0){10}\makebox(0,0)[lc]{$x$}}

\put(36,1){\vector(-1,0){14}}\put(36,1){\vector(1,0){14}}
\put(36,2){\makebox(0,0)[cb]{$L=c\tau$}}

\end{picture}

(b)

\caption{Reflection from a moving mirror (a) and a Fabry-Perot cavity
(b)}\label{fig:quant}

\end{center}

\end{figure}
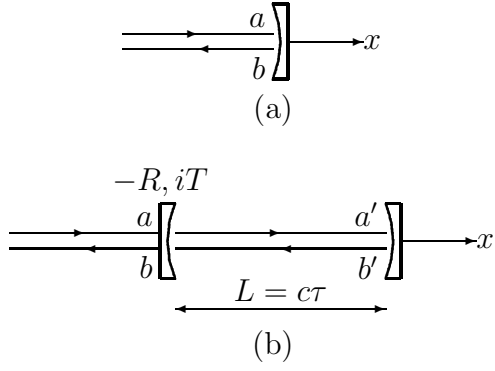

Consider first a single moving mirror with reflection $R=-1$. Let $x(t)$ be
the current position of the mirror and $a,b$ are the annihilation operators
corresponding to the incident and reflected beams (see
Fig.\,\ref{fig:quant}(a)).

Suppose that the incident beam can be presented as a sum of a strong classical
monochromatic wave with frequency $\omega_o$ and small quantum term, which we
refer below to as ``zeroth approximation'' and ``first approximation'',
correspondingly:

\begin{equation}\label{mirr_a}
  \hat E_a(t) = E_0(\omega_o)Ae^{-i\omega_o t}
    + \intOinfty E_0(\omega)\hat a(\omega)e^{-i\omega t}\,\frac{d\omega}{2\pi}
    + \hc \,.
\end{equation}
Here

\begin{equation}
  E_0(\omega) = \sqrt{\frac{2\pi\hbar\omega}{{\cal A}c}}
\end{equation}
is the normalization factor, ${\cal A}$ is the cross-section area of the beam,
and $\hc$ stands for ``Hermitian conjugate''. This value of the normalization
factor corresponds to the commutator

\begin{equation}
  [\hat a(\omega),\hat a^+(\omega')] = 2\pi\delta(\omega-\omega') \,.
\end{equation}

Neglecting relativistic terms proportional to $1/c^2$ the reflected beam can
be presented in the following form:

\begin{equation}\label{MM_prec}
  \hat E_b(t) \approx -\frac{1-\hat{\dot x}(t)/c}{1+\hat{\dot x}(t)/c}\,
    \hat E_a(t-2x(t)/c)
  \approx \hat E_a(t) - \frac{1}{c}\,\frac{d\hat x(t)\hat E_a(t)}{dt} \,.
\end{equation}
Taking into account formula (\ref{mirr_a}) expression (\ref{MM_prec}) can be
rewritten as:

\begin{equation}\label{mirr_b}
  \hat E_b(t) = E_0(\omega_o)Be^{-i\omega_o t}
    + \intOinfty E_0(\omega)\hat b(\omega)e^{-i\omega t}\,\frac{d\omega}{2\pi}
    + \hc \,,
\end{equation}
where

\begin{gather}
  B = -A \,, \nonumber \\
  \hat b(\omega) = -\hat a(\omega)
    - 2i\varkappa(\omega)A\hat x(\omega_o-\omega) \,,
    \label{MM} \\
  \varkappa(\omega) = \frac{\sqrt{\omega_o\omega}}{c} \,,
\end{gather}
and

\begin{equation}
  \hat x(\Omega) = \intinfty\hat x(t)e^{-i\Omega t}\,\frac{d\Omega}{2\pi}
\end{equation}
is the spectrum of the $\hat x(t)$.

\subsubsection{Reflection from a Fabry-Perot cavity}\label{sec:FP}

Suppose then that the second mirror with the reflection $-R$ and transmittance
$iT$ ($R^2+T^2=1$) is added to the first one forming a Fabry-Perot cavity with
the length $L=c\tau_1$ [see Fig.\,\ref{fig:quant}(b)].

Using the same representation (\ref{mirr_a}) as in the previous subsubsection,
equation for the field amplitudes can be written as the following:

\begin{align}
  B &= -RA + iTB'e^{i\omega_o\tau_1}   \,, \nonumber \\
  A' &= iTAe^{i\omega_o\tau_1}  - RB'e^{2i\omega_o\tau_1} \,, \nonumber \\
  B' &= -A' \,,
\end{align}
\begin{align}
  \hat b(\omega) &= -R\hat a(\omega) + iT\hat b'(\omega)e^{i\omega\tau_1}\,,
    \nonumber \\
  \hat a'(\omega) &=
    iT\hat a(\omega)e^{i\omega\tau_1} - R\hat b'(\omega)e^{2i\omega\tau_1}\,,
    \nonumber \\
  \hat b'(\omega) &= -\hat a'(\omega)
    - 2i\varkappa(\omega)A'\hat x(\omega_o-\omega)\,.
\end{align}
Solution of these equations is equal to:

\begin{align}
  B &= {\cal R}(\omega_o)A \,, \nonumber \\
  A' &= -B' = {\cal T}(\omega_o)A\,, \label{FP_0}
\end{align}
\begin{align}
  \hat b(\omega) &= {\cal R}(\omega)\hat a(\omega)
    + \varkappa(\omega)A'{\cal S}(\omega)\hat x(\omega_o-\omega) \,,
      \label{FP_1(a)}\\
  \hat a'(\omega) - \hat b'(\omega) &= 2{\cal T}(\omega)\hat a(\omega)
    + \varkappa(\omega)A'{\cal W}(\omega)\hat x(\omega_o-\omega) \,,
      \label{FP_1(b)}
\end{align}
where

\begin{align}\label{FP_1p}
  {\cal R}(\omega) &= \frac{{\cal L}^*(\omega)}{{\cal L}(\omega)}  \,, &
  {\cal S}(\omega) &= \frac{2\sqrt{\varGamma}}{{\cal L}(\omega)} \,,
  \nonumber \\
  {\cal T}(\omega) &= \frac{i\sqrt{\varGamma}}{{\cal L}(\omega)} \,, &
  {\cal W}(\omega) &= 2i\,\frac{{\cal L}'(\omega)}{{\cal L}(\omega)} \,,
\end{align}
and

\begin{align}
  {\cal L}(\omega) &= \varGamma\cos\omega\tau_1-i\sin\omega\tau_1 \,,\nonumber \\
  {\cal L}'(\omega) &= \cos\omega\tau_1-i\varGamma\sin\omega\tau_1 \,,\nonumber\\
  \varGamma &= \frac{1-R}{1+R}\,.
\end{align}

These formulas can be simplified if the pumping frequency is equal to one of
the eigenfrequencies of the Fabry-Perot cavity, $\omega_o\tau_1 = 2\pi n$, and
the observation frequency is relatively small, $\Omega\tau_1\ll 1$. In this case

\begin{align}
  {\cal R}(\omega_o+\Omega)
    &\approx \frac{\gamma+i\Omega}{\gamma-i\Omega}\,, &
  {\cal S}(\omega_o+\Omega)
    &\approx \frac{2\sqrt{\gamma/\tau_1}}{\gamma-i\Omega}\,,\nonumber \\
  {\cal T}(\omega_o+\Omega)
    &\approx \frac{\sqrt{\gamma/\tau_1}}{\gamma-i\Omega}\,, &
  {\cal W}(\omega_o+\Omega)
    &\approx \frac{2i}{\tau_1}\,\frac{1}{\gamma-i\Omega} \,,
\end{align}
where $\gamma = \varGamma/\tau_1$ is the half-bandwidth of the cavity.

\subsubsection{Pondermotive force}

The formula (\ref{FP_1(b)}) allows to calculate the pondermotive force acting
on the moving mirror in a Fabry-Perot cavity. This force is equal to

\begin{equation}
  F(t) = \frac{{\cal A}}{4\pi}
    \left[\overline{E_{a'}^2(t)} + \overline{E_{b'}^2(t)}\right] \,,
\end{equation}
where overline means averaging over time larger than the period of the optical
oscillations  $2\pi/\omega_o$. This force can be presented as a sum of the
large constant force

\begin{equation}
  F_0 = \frac{2W}{c} = \frac{2w}{c\varGamma} \,,
\end{equation}
where $W=|A'|^2/c$ is the mean value of the power circulating in the cavity
and $w=|A|^2/c$ is the mean power of the incident beam, and the fluctuating
force

\begin{multline}\label{FP_pond}
  \hat F(t) = \hbar{A'}^*\intOinfty\varkappa(\omega)
    \bigl[\hat a'(\omega)-\hat b'(\omega)\bigr]e^{i(\omega_o-\omega)t}\,
    \frac{d\omega}{2\pi} + \hc \\
  = \hbar\intOinfty\left[
      2\varkappa(\omega){A'}^*{\cal T}(\omega)\hat a(\omega)
      + \varkappa^2(\omega)|A'|^2{\cal W(\omega)}\hat x(\omega_o-\omega)
    \right]e^{i(\omega_o-\omega)t}\,\frac{d\omega}{2\pi} + \hc \,.
\end{multline}

\subsection{The optical speedmeter}

\begin{figure}[t]

\begin{center}

\begin{picture}(140,140)

\put(0,0){\usebox{\beamsplMain}}
\put(35,35){\usebox{\beamsplSENW}}\put(35,35){\makebox(0,0)[rt]{\sf PBS}}
\put(100,35){\usebox{\FabryH}}
\put(35,100){\usebox{\FabryV}}

\thicklines

\put(70,38){\framebox(10,4)}\put(75,37){\makebox(0,0)[ct]{\sf FR1}}
\put(38,70){\framebox(4,10)}\put(37,75){\makebox(0,0)[rc]{\sf FR2}}

\thinlines

\drawline(40,70)(40,40)(70,40)

\put(55,41){\vector(1,0){10}}
\put(55,43){\shortstack{$a_2(0^\circ)$\\$g_1(90^\circ)$}}
\put(65,39){\vector(-1,0){10}}
\put(55,30){\shortstack{$b_2(0^\circ)$\\$g_2(90^\circ)$}}
\drawline(80,40)(100,40)
\put(85,41){\vector(1,0){10}}
\put(84,43){\shortstack{$a_2(45^\circ)$\\$g_1(135^\circ)$}}
\put(95,39){\vector(-1,0){10}}
\put(84,30){\shortstack{$b_2(135^\circ)$\\$g_2(45^\circ)$}}

\put(39,55){\vector(0,1){10}}
\put(26,55){\shortstack{$a_1(0^\circ)$\\$g_2(90^\circ)$}}
\put(41,65){\vector(0,-1){10}}
\put(42,55){\shortstack{$b_1(0^\circ)$\\$g_1(90^\circ)$}}
\drawline(40,80)(40,100)
\put(39,85){\vector(0,1){10}}
\put(24,85){\shortstack{$a_1(45^\circ)$\\$g_2(135^\circ)$}}
\put(41,95){\vector(0,-1){10}}
\put(42,85){\shortstack{$b_1(135^\circ)$\\$g_1(45^\circ)$}}

\end{picture}

\end{center}

\caption{The optical speedmeter}\label{fig:thescheme}

\end{figure}
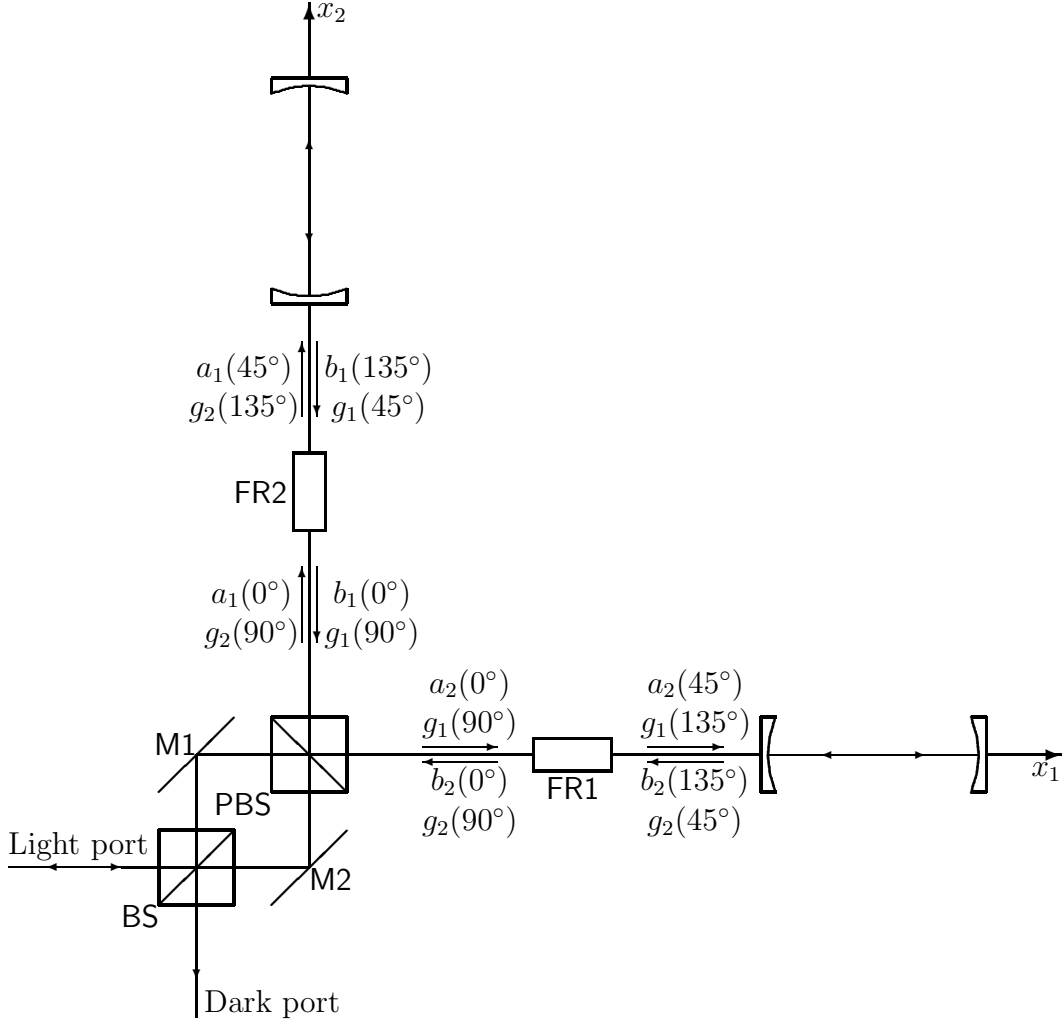

\subsubsection{General remarks and notations}

The following notations will be used in this subsection (see
Fig.\,\ref{fig:thescheme}):

$a_1$ is the amplitude of the beam which goes from the main beam splitter {\sf
BS} to the right directions, is reflected from the mirror {\sf M2}, passes
through the polarization beam splitter {\sf PBS}, and enters the upper
Fabry-Perot cavity with $45^\circ$-polarization;

$g_1$ is the amplitude of this beam after the reflection from the upper
Fabry-Perot cavity; it is reflected from the {\sf PBS} and enters into the
right Fabry-Perot cavity with $135^\circ$-polarization;

$b_2$ is the amplitude of this beam after the reflection from the right
Fabry-Perot cavity; it passes through the {\sf PBS} and returns to the {\sf
BS} from the upper direction;

$a_2$, $g_2$ and $b_1$ describe the beam which goes in the opposite direction
in the similar way;

$a_1'(45^\circ)$ and $a_1'(135^\circ)$ are the amplitudes pumped in the first
(right) cavity by the beams $a_2$ and $g_1$, correspondingly;

$a_2'(45^\circ)$ and $a_2'(135^\circ)$ are the amplitudes pumped in the second
(upper) cavity by the beams $a_1$ and $g_2$, correspondingly;

$a_3$ corresponds to the beam which goes from the {\sf BS} to the left
direction (to the pumping laser), and $b_4$ --- to the beam which goes from
the laser to the {\sf BS};

$a_4$ corresponds to the beam which goes from the {\sf BS} to the down
direction (to the detector), and $b_4$ --- to the beam which goes from
the detector to the {\sf BS}.

The main beam splitter {\sf BS} is described by the following equations:

\begin{align}
  \hat a_1 &= -\frac{\hat b_3+i\hat b_4}{\sqrt 2} \,, &
  \hat a_2 &= -\frac{\hat b_3-i\hat b_4}{\sqrt 2} \,, \nonumber \\
  \hat a_3 &= -\frac{\hat b_1+\hat b_2}{\sqrt 2} \,, &
  \hat a_4 &= -i\frac{\hat b_1-\hat b_2}{\sqrt 2} \,. \label{SM_beamspl}
\end{align}
It is presumed here that there is the additional phase shift $\pi/2$ in the
right port ``1'' of the beam splitter, which is necessary in order to provide
the dark fringe regime (exactly as in the standard topology).

\subsubsection{Field amplitudes}

In the zeroth approximation in addition to the equations (\ref{SM_beamspl})
the following equations can be written:

\begin{align}
  B_{1,2} &= G_{2,1} \,, & G_{1,2} &= A_{1,2} \,, \label{SM_0r}  \\
  A_{1,2}'(45^\circ) &= \frac{iA_{2,1}}{\sqrt\varGamma} \,, &
  A_{1,2}'(135^\circ) &= \frac{iG_{1,2}}{\sqrt\varGamma} \,. \label{SM_0i}
\end{align}
Solving this set of equations and taking into account that $B_4=0$ (there is
no pumping in the dark port), we obtain that

\begin{gather}
  A_3 = B_3 \,, \qquad A_4 = 0 \,, \nonumber \\
  A_1'(45^\circ) = A_1'(135^\circ) = A_2'(45^\circ) = A_2'(135^\circ)
    \equiv A' = -\frac{iB_3}{\sqrt{2\varGamma}} \label{SM_0} \,.
\end{gather}

In the first approximation the following equation can be obtained
from the formula (\ref{FP_1(a)}):

\begin{align}
  \hat b_{1,2}(\omega) &= {\cal R}(\omega)\hat g_{2,1}(\omega)
    + \varkappa(\omega)A_{2,1}'(135^\circ){\cal S}(\omega)
        \hat x_{2,1}(\omega_o-\omega) \,, \nonumber \\
  \hat g_{1,2}(\omega) &= {\cal R}(\omega)\hat a_{1,2}(\omega)
    + \varkappa(\omega)A_{2,1}'(45^\circ){\cal S}(\omega)
        \hat x_{2,1}(\omega_o-\omega) \label{SM_1r} \,.
\end{align}
Solution of the equations (\ref{SM_beamspl},\ref{SM_1r}) is equal to:

\begin{align}
  \hat b_{1,2}(\omega) &= {\cal R}^2(\omega)\hat a_{2,1}(\omega)
   + \varkappa(\omega)A'{\cal S}(\omega)\bigl[
       {\cal R}(\omega)\hat x_{1,2}(\omega_o-\omega)
       + \hat x_{2,1}(\omega_o-\omega)
     \bigr] \,, \label{SM_1b} \\
  \hat g_{1,2}(\omega) &= {\cal R}(\omega)\hat a_{1,2}(\omega)
    + \varkappa(\omega)A'{\cal S}(\omega)\hat x_{2,1}(\omega_o-\omega)\,,
    \label{SM_1g} \\
  \hat a_4(\omega) &= {\cal R}^2(\omega)\hat b_4(\omega)
    + i\sqrt{2}\varkappa(\omega)A'{\cal S}(\omega)
        \bigl[1-{\cal R}(\omega)\bigr]\hat x_-(\omega) \,. \label{SM_out}
\end{align}

\subsubsection{Measurement noise}

Using the formula (\ref{SM_out}), it can be shown that the output signal of
the homodyne detector is proportional to\footnote{There is no factor
$E_0(\omega)\propto\sqrt{\omega}$ in the next formula because the photocurrent
is proportional not to the optical power but to the quanta flux.}

\begin{equation}
  \hat{\cal I}_{\rm detect}(t) = \intOinfty\hat a_4(\omega)
    e^{i(\omega_o-\omega)t+i\phi_{\rm LO}}\,\frac{d\omega}{2\pi} + \hc
  = \hat{\cal I}_{\rm detect}(t) + \hat{\cal I}_{\rm noise}(t) \,,
\end{equation}
where $\phi_{\rm LO}$ is the phase of the local oscillator,

\begin{equation}\label{SM_Enoise}
  \hat{\cal I}_{\rm noise}(t) = \intOinfty{\cal R}^2(\omega)\hat b_4(\omega)
    e^{i(\omega_o-\omega)t+i\phi_{\rm LO}}\,\frac{d\omega}{2\pi} + \hc
\end{equation}
is the noise,

\begin{equation}
  \hat{\cal I}_{\rm signal}(t)
  = \intinfty k(\Omega)\hat x(\Omega)e^{i\Omega t}\,\frac{d\Omega}{2\pi}
\end{equation}
is the signal,

\begin{equation}
  k(\Omega) = \frac{
    4i\sqrt{\omega_o}|B_3|\sin\Omega\tau_1\bigl[
      \sqrt{\omega_o-\Omega}\,e^{i\phi} + \sqrt{\omega_o+\Omega}\,e^{-i\phi}
    \bigr]
  }{c(\varGamma\cos\Omega\tau_1+i\sin\Omega\tau_1)^2}
\end{equation}
is the transmission factor and $\phi = \phi_\mathrm{LO} + \arg B_3$ (note the
factor $\sin\Omega\tau_1$).

If the pumping beam is in the coherent quantum state, then the spectral
density of the noise $\hat {\cal I}_{\rm noise}(t)$ is equal to

\begin{equation}
  S_{\cal I}(\Omega) = 1 \,,
\end{equation}
In this case the spectral density of the measurement noise is equal to

\begin{equation}\label{S_x}
  S_x(\Omega) = \frac{S_{\cal I}(\Omega)}{|k(\Omega)|^2}
  = \frac{\hbar c^2(\varGamma^2\cos^2\Omega\tau_1 + \sin^2\Omega\tau_1)^2}{
      16w\sin^2\Omega\tau_1\bigl|
        \sqrt{\omega_o+\Omega}\,e^{i\phi} + \sqrt{\omega_o-\Omega}\,e^{-i\phi}
      \bigr|^2
    }\,,
\end{equation}
where $w=\hbar\omega_o|B_3|^2$ is the pumping power.

\subsubsection{Back-action}

Difference of the pondermotive (back action) forces which act on the distant
mirrors of the Fabry-Perot cavities, is equal to

\begin{equation}
  \hat F_-(t) = \hat F_1^{45^\circ}(t) + \hat F_1^{135^\circ}(t)
    - \hat F_2^{45^\circ}(t) - \hat F_2^{135^\circ}(t) \,,
\end{equation}
where [see the formulas (\ref{FP_pond}, \ref{SM_0}, \ref{SM_1b}, \ref{SM_1g})]

\begin{align}
  \hat F_{1,2}^{45^\circ}(t) &= \hbar\intOinfty\left[
    2\varkappa(\omega){A'}^*{\cal T}(\omega)\hat a_{2,1}(\omega)
    + \varkappa^2(\omega)|A'|^2{\cal W(\omega)}\hat x_{1,2}(\omega_o-\omega)
  \right]e^{i(\omega_o-\omega)t}\,\frac{d\omega}{2\pi} + \hc \,, \nonumber \\
  \hat F_{1,2}^{135^\circ}(t) &= \hbar\intOinfty\left[
    2\varkappa(\omega){A'}^*{\cal T}(\omega)\hat g_{1,2}(\omega)
    + \varkappa^2(\omega)|A'|^2{\cal W(\omega)}\hat x_{1,2}(\omega_o-\omega)
  \right]e^{i(\omega_o-\omega)t}\,\frac{d\omega}{2\pi} + \hc \,.
\end{align}
Taking into account formulas (\ref{SM_0}, \ref{SM_1b}, \ref{SM_1g}), we obtain
that

\begin{equation}
  \hat F_-(t) = \hat F_{\rm fluct}(t) + \hat F_{\rm dyn}(t)\,,
\end{equation}
where

\begin{equation}\label{SM_Ffluct}
  \hat F_{\rm fluct}(t) = -2\frac{\hbar B_3^*}{\sqrt{\varGamma}}\intOinfty
    \varkappa(\omega){\cal T}(\omega)\bigl[1-{\cal R}(\omega)\bigr]
    \hat b_4(\omega)e^{i(\omega_o-\omega)t}\,\frac{d\omega}{2\pi} + \hc
\end{equation}
is the fluctuational back action and

\begin{multline}
  \hat F_-(t) = 2\frac{\hbar|B_3|^2}{\varGamma}\intOinfty\varkappa^2(\omega)
    \bigl[{\cal W}(\omega)-{\cal S}(\omega){\cal T}(\omega)\bigr]
    \hat x_-(\omega_o-\omega)e^{i(\omega_o-\omega)t}\,\frac{d\omega}{2\pi}
  + \hc \\
  = - \intinfty K(\Omega)\hat x_-(\Omega) \,,
\end{multline}
is the dynamic back action. The factor $K(\Omega)$ (the pondermotive rigidity)
is equal to

\begin{equation}
  K(\Omega) = -\frac{8w\Omega}{c^2\varGamma}\,
    \frac{(1+\varGamma^2)\cos\Omega\tau_1\sin\Omega\tau_1+i\Omega\sin^2\Omega\tau_1}
      {(\varGamma\cos\Omega\tau_1+i\sin\Omega\tau_1)^2}
  \approx (\text{if $\Omega\tau_1\ll 1$}) \approx
    \frac{8w}{L^2\gamma}\,\frac{(i\Omega)^2}{(\gamma+i\Omega)^2} \,.
\end{equation}
This is a very small value: if, for example, $w=1{\rm KWt}$, $L=4{\rm Km}$,
and $\gamma\sim\Omega\sim 10^3{\rm s}^{-1}$ then $K\sim 10^{-3}\,{\rm
dyn/cm^2}$. Therefore, we can conclude that the dynamic back action is
practically absent in this scheme.

If the pumping field is in the coherent quantum state then the spectral
density of the fluctuational force (\ref{SM_Ffluct}) is equal to

\begin{equation}\label{S_F}
  S_F(\Omega) = \frac{16\hbar\omega_o w}{c^2}\,
    \frac{\sin^2\Omega\tau_1}
      {(\varGamma^2\cos^2\Omega\tau_1+\sin^2\Omega\tau_1)^2} \,,
\end{equation}
and the cross-correlation spectral density of the noises $\hat{\cal I}_{\rm
noise}$ and $\hat F_{\rm fluct}(t)$ is equal to

\begin{equation}
  S_{{\cal I}F}(\Omega) = \frac{
    2\hbar\sqrt{\omega_o}|B_3|\sin\Omega\tau_1\bigl[
      \sqrt{\omega_o+\Omega}\,e^{i\phi} - \sqrt{\omega_o-\Omega}\,e^{-i\phi}
    \bigr]
  }{c(\varGamma\cos\Omega\tau_1 - i\sin\Omega\tau_1)^2} \,.
\end{equation}
Therefore, cross-correlation spectral density for the measurement noise and
the back-action noise is equal to

\begin{equation}\label{S_xF}
  S_{xF}(\Omega) = \frac{S_{{\cal N}F}(\Omega)}{k^*(\Omega)}
  = \frac{i\hbar}{2}\,\frac{
      \sqrt{\omega_o+\Omega}\,e^{i\phi} - \sqrt{\omega_o-\Omega}\,e^{-i\phi}
    }{
      \sqrt{\omega_o+\Omega}\,e^{i\phi} + \sqrt{\omega_o-\Omega}\,e^{-i\phi}
    } \,.
\end{equation}
It have to be noted, that

\begin{equation}
  S_x(\Omega)S_F(\Omega) - |S_{xF}(\Omega)|^2 = \frac{\hbar^2}{4} \,.
\end{equation}

\subsubsection{The spectral densities}

It is easy to note that the formulas (\ref{S_x}, \ref{S_F}, \ref{S_xF}) can be
simplified. Really, the observation frequency $\Omega\lesssim 10^3\,{\rm
s}^{-1}$ is much smaller than the pumping frequency $\omega_o \sim
10^{15}\,{\rm s}^{-1}$. Therefore, the formulas (\ref{S_x}, \ref{S_xF}) can be
rewritten in the following form:

\begin{align}
  S_x(\Omega) &= \frac{\hbar c^2}{64\omega_ow\cos^2\phi}\,
    \frac{(\varGamma^2\cos^2\Omega\tau_1 + \sin^2\Omega\tau_1)^2}
      {\sin^2\Omega\tau_1} \,, \label{S_x_1} \\
  S_{xF}(\Omega) &= -\frac{\hbar}{2}\,\tan\phi \label{S_xF_1} \,.
\end{align}
Then, in the real gravitational-wave antennae, $\tau_1\lesssim 10^{-5}\,s$,
and hence $\Omega\tau_1$ is also a small parameter. This allows to further
simplify the formulas (\ref{S_F}, \ref{S_xF_1}):

\begin{align}
  S_x(\Omega) &= \frac{\hbar L^2}{64\omega_ow\cos^2\phi}\,
    \frac{(\gamma^2 + \Omega^2)^2}{\Omega^2} \label{S_x_2} \,, \\
  S_F(\Omega) &= \frac{16\hbar\omega_o w}{L^2}\,
    \frac{\Omega^2}{(\gamma^2 + \Omega^2)^2}  \label{S_F_2} \,,
\end{align}
where $\gamma = \varGamma/\tau_1$. Therefore, the spectral densities $S_v,
S_p$ and $S_{vp}$ can be presented in the following form:

\begin{align}
  S_v(\Omega) &\equiv \Omega^2S_x(\Omega)
    = \frac{\hbar L^2(\gamma^2 + \Omega^2)^2}{64\omega_ow\cos^2\phi}\,,
    \label{S_v} \\
  S_p(\Omega) &\equiv \frac{S_F(\Omega)}{\Omega^2}
    = \frac{16\hbar\omega_o w}{L^2(\gamma^2 + \Omega^2)^2}\,, \label{S_p} \\
  S_{vp}(\Omega) &\equiv -S_{xF}(\Omega)
    = \frac{\hbar}{2}\,\tan\phi \label{S_vp} \,.
\end{align}


\end{document}